%% file: main_arxiv.tex
\documentclass[12pt]{article}
\usepackage{fancyhdr}
\usepackage{algpseudocode}
\usepackage{amsmath, amssymb, amsthm}
\usepackage{float, fullpage, graphicx, multirow,parskip, subcaption, setspace}
\usepackage{comment}
\usepackage{url}
\usepackage{enumitem}
\usepackage{tikz}
\usepackage{bm, bbm}
\usetikzlibrary{shapes.geometric, positioning}
\usetikzlibrary{quotes, angles}
\usepackage{rotating}
\usepackage{hyperref}
\usepackage{natbib}
\usepackage{placeins}
\usepackage{algorithm2e}
\usepackage{geometry}
\RestyleAlgo{ruled}
\usepackage{lineno}

\usepackage{rotating}
\usepackage{booktabs} 
\usepackage{longtable}
\usepackage{tabularx,colortbl}

\definecolor{SkyBlue}{RGB}{14, 118, 188}
\definecolor{BrightRed}{RGB}{223,82, 78}

\newcommand{\cD}{\mathcal{D}}
\newcommand{\cH}{\mathcal{H}}

\hypersetup{pdfborder = {0 0 0.5 [3 3]}, colorlinks = true, linkcolor = BrightRed, citecolor = SkyBlue}

\def\keywordname{{\bfseries \emph Keywords}}%
\def\keywords#1{\par\addvspace\medskipamount{\rightskip=0pt plus1cm
\def\and{\ifhmode\unskip\nobreak\fi\ $\cdot$
}\noindent\keywordname\enspace\ignorespaces#1\par}}

\title{Geometric theory on large-scale and local determination of density dependence of a recovering large carnivore population}

\author{Yunyi Shen$^{1, *}$ \and Erik R. Olson$^2$ \and Timothy R. Van Deelen$^3$ }

\date{
\small $^1$ Laboratory for Information and Decision Systems, Department of Electrical Engineering and Computer Science,  Massachusetts Institute of Technology.\\
$^{2}$ Department of Environmental Sciences, Northland College.\\
$^3$ Department of Forest and Wildlife Ecology, University of Wisconsin--Madison.\\
$^*$ Corresponding author: yshen99@mit.edu
}

\begin{document}
\begin{spacing}{1.5}

\maketitle


\begin{abstract}
    Density-dependent population growth is a feature of large carnivores like wolves (\textit{Canis lupus}), with mechanisms typically attributed to resource (e.g. prey) limitation. Such mechanisms are local phenomena and rely on individuals having access to information, such as prey availability at their location. Using over four decades of wolf population and range expansion data from Wisconsin (USA) wolves, we found that the population not only exhibited density dependence locally but also at landscape scale. Superficially, one may consider space as yet another limiting resource to explain landscape-scale density dependence. However, this view poses an information puzzle: most individuals do not have access to global information such as range-wide habitat availability as they would for local prey availability. How would the population ``know'' when to slow their range expansion? To understand observed large-scale spatial density dependence, we propose a reaction-diffusion model, first introduced by Fisher and Kolmogorov, with a ``travelling wave'' solution, wherein the population expands from a core range that quickly achieves  local carrying capacity. Early-stage acceleration and later-stage deceleration of population growth can be explained by early elongation of an expanding frontier and a later collision of the expanding frontier with a habitat boundary. Such a process does not require individuals to have global density information. We illustrate our proposal with simulations and spatial visualizations of wolf recolonization in the western Great Lakes region over time relative to habitat suitability. We further synthesize previous studies on wolf habitat selection in the western Great Lakes region and argue that the habitat boundary appeared to be driven by spatial variation in mortality, likely associated with human use of the landscape.
\end{abstract}

\section{Introduction}
\label{sec:introduction}
\input{tex/introduction}

\section{Material and Methods}
\label{sec:method}
\input{tex/method}

\section{Results}
\label{sec:results}
\input{tex/results}

\section{Large scale density dependence emerge from a reaction-diffusion model}
\label{sec:theory}
\input{tex/theory}

\section{Local density dependence and carrying capacity landscape}
\label{sec:niche}
\input{tex/niche_theory}

\section{Discussion}
\label{sec:discussion}
\input{tex/discussion}

\section*{Acknowledgement}
\input{tex/acknowledgement}

\FloatBarrier
\bibliography{references.bib}

\newpage
\FloatBarrier
\section*{Appendix}
\label{sec:appendix}
\input{tex/appendix.tex}

\end{spacing}
\end{document}

%% file: tex/introduction.tex

Density dependence occurs when a population's per capita growth rate is influenced by its density through density feedback mechanisms \citep{sinclair1996density, herrando2012density}. It is a pervasive feature of animal population dynamics \citep{sinclair1996density, sibly2005regulation, brook2006strength}.
In the interest of standard language across a taxonomically and topically diverse literature in ecology \citep{herrando2012density}, an ensemble density feedback is the realization, on per capita growth rate, of one or more component density feedbacks acting on demographic components of growth (e.g. reproduction). Demonstration of an ensemble density feedback is evidence that at least one component density feedback is operating although the reverse is not necessarily true because of potentially offsetting component effects elsewhere \citep{stephens1999consequences, berec2007multiple}. However, some offsetting effects such as compensatory behavior among cause-specific mortalities are themselves evidence of density dependence \citep{peron2013compensation}. 
Component density feedbacks usually have direct exogenous effects on a demographic rate such as when per capita food shortages with increasing density lead to increase starvation mortality or reduced reproduction \citep{solberg2001effects}. Such direct effects can also, at times, be mediated by indirect effects such as encounters with conspecifics or behavioral cues like scent marking and vocalizations \citep{peters1975scent, harrington2003wolf}. For instance, wolves exhibit density dependence through territorial behavior thereby creating a carrying capacity in local density \citep{o2017spatially,o2019territorial}. Among predatory mammals, large size is associated with selection for endogenous socio-behavioral controls on growth such as female reproductive suppression, female territoriality, and cooperative behaviors that regulate per capita growth and mitigate direct links between density and discrete vital rates \citep{wallach2015apex}. Understanding how these mechanisms work is essential for understanding and predicting equilibrium densities \citep{van2009growth}, the potential for prey regulation \citep{messier1994ungulate}, trophic dynamics \citep{wallach2015apex} and population responses to perturbation \citep{stenglein2018compensatory}.
Recently recovered wolf (\textit{Canis lupus}) populations (USA) show negative ensemble density feedback. This is evident in the numerical and spatial patterns of population growth \citep{van2009growth,stenglein2016demographic,o2019territorial} and in components of growth like survival \citep{cubaynes2014density, o2017spatially, stenglein2018compensatory}, reproduction \citep{stahler2013adaptive,stenglein2015numbers} and compensatory behavior in cause-specific mortalities \citep{hochard2014gray,stenglein2018compensatory}.
Evidence for a socio-behavioral mediated depensatory feedback in wolf population growth at low density comes from modeling reduced mate-finding success due to limited perception range at the margins of expanding populations \citep{hurford2006spatially, stenglein2015individual}.  However, \citet{wallach2015apex}'s predictions of a compensatory (negative) feedback mediated by socio-behavioral effects at high-density prompt an important life history question: How do territorial animals with limited perception ranges \citep{stenglein2015individual} ``know'' that their population's global range is beginning to saturate such that socio-behavioral mitigators of growth can operate? Do they need to know? Or, does local density interact with the population's expanding wave-front such that global information isn't needed? Recovery of wolf population in Wisconsin and the western Great Lakes regions provides a unique opportunity to study these questions since both occupied range and population trend were continuously monitored for more than four decades -- from extirpation to recovery. 
Our goals were 1) to examine these questions by quantifying relationships between local and global densities, spatial behavior, and population growth using data from a recovered wolf population in Wisconsin, USA \citep{wiedenhoeft2020wisconsin}, 2) to propose and examine a classic reaction-diffusion model \citep{fisher1937wave, kolmogorovstudy} as a way to describe the interplay between local density and range boundaries for an expanding population. 3) to examine the nature of habitat boundaries and determination of local carrying capacity for wolves throughout the western Great Lake regions. 


%% file: tex/method.tex
\subsection{Wolf demographic and range data}

Wolf demography, pack size (number of wolves in each pack), and territory data for the state of Wisconsin, USA were obtained from the Wisconsin Department of Natural Resources. We used winter minimum counts for wolves from 1979-2020 as our population metric \citep[e.g.][]{wiedenhoeft2020wisconsin}. We estimated wolf range in Wisconsin for each year by calculating area in km$^2$ of a simplified 5-km buffer around all wolf pack territories present in a given year, using ArcMap \citep[version 10,][]{arcmap}. We used 5 km buffers because they have been used as an indicator for extraterritorial movement for very-high frequency (VHF) radio-collared wolves \citep{messier1985solitary,fuller1989population}. A 5-km buffer also minimizes any small-scale errors associated with the representation of wolf pack territories on an annual basis. To include the western Great Lakes regions, namely the additional states of Michigan and Minnesota, USA, we obtained wolf demographic and range data directly from the corresponding Departments of Natural Resources \citep{erb2015minnesota, midnr}.

\subsection{Landscape density dependence}

To obtain a landscape level understanding of wolf population demographics, we concentrate solely on demography. Our analysis involved inferring an overall carrying capacity in terms of both the population range and the total population count. To achieve this, we employed a non-linear regression model that incorporated a logistic growth mean along with normal and Poisson observations, respectively. The objective of this analysis was to test the hypothesis of a finite carrying capacity ($K$) for range and population. To assess this, we conducted a likelihood ratio test comparing the null hypothesis $H_0: 1/K=0$ against the alternative $H_1:1/K\ne 0$. Effectively, this test allows us to determine whether the wolf population or range exhibited logistic growth or not.

To ensure analysis of logistic growth pattern remained unaffected by legal harvests during 2012-2014, we first used data from 1980 to 2011. Additionally, we conducted a sensitivity analysis on our test results. This involved performing the same analysis using two additional datasets: one comprising the entire dataset, and the other excluding only the years 2012-2014 (i.e. 1980-2011 and 2015-2020).  

One can view harvest effects as essentially resetting the population back to an earlier initial value in the timeseries. These ``reset'' dynamics were used to further validate our model.  To do so, we  fitted logistic curves to predict population dynamics from 2015 to 2020. To accomplish this, we set post-harvest years (i.e., 2015-2020) back in time, aligning them to begin in 2011, prior to the initiation of the legal wolf harvest in the state and aligning most closely with the 2015 post-harvest population. Finally, we compared our estimated carrying capacity with values found in the existing literature, which were derived from species distribution models  \citep{mladenoff2009change, gantchoff2022distribution}, total occupied wolf range \citep{stauffer2021scaling}, and phenomenological growth models \citep{van2009growth}.

\subsection{Patterns in local density}
\label{sec:average_local_den}
Acknowledging the spatial structure of wolf recovery, we investigated whether local density remained stable across occupied wolf range occupied. We used linear regression between population size and range, without intercept. Our expectation was that if local densities remained relatively constant across wolf range, then a large portion of the variability in population size would be explained by wolf range alone. To further assess the stability of local wolf densities between 2000 and 2020, we employed both linear regression with time and Kwiatkowski-Phillips-Schmidt-Shin (KPSS) test \citep{shin1992kpss}. The KPSS test directly examined whether local density (population divided by range) during this period exhibited stationarity, indicating stability over time. 

The relationship obtained through linear regression of population size and range offers an alternative approach to estimating landscape-level carrying capacity. This method does not rely on demographic models but instead uses total habitat range estimates reported in the literature \citep{mladenoff2009change, gantchoff2022distribution, stauffer2021scaling}.

To assess consistency between estimates, we compared carrying capacity derived from  such habitat-based estimates with carrying capacity inferred from demographic analysis.

\subsection{Pack size during expansion}
Examining whether average pack or territory size changes during range expansion would give insights into biological mechanisms behind expansion, i.e., whether it is driven by changes in the number of packs, wolf pack size, or wolf pack territory size. We modeled variation in average pack and territory size using linear regression of population size (and range) relative to number of packs.

\subsection{Local wolf density and prey density}

We collected density estimates of white tailed deer (\textit{Odocoileus virginianus}),the main prey of wolves \citep{thompson1952travel} in our study area. We obtained deer density information at the deer management unit-scale from the Wisconsin Department of Natural Resources. We then generated area-weighted mean deer densities for each wolf management zone for 1980-2012. We also calculated the empirical local density of wolves for each wolf management zone in each year by dividing number of wolves in a zone by the area of wolf range in that zone in the same time frame. To test if deer density explained variation in local wolf density, we performed a linear regression between estimated deer density and wolf density at zone-year level.

\subsection{Mapping expansion of wolves across spatial scales}
To analyze wolf range expansion across spatial scales, we examined wolf range and population dynamics at state-wide and management zone scales. To estimate number of wolves within each Wisconsin wolf management zone each year, we calculated the proportion of a wolf pack territory within each zone and then multiplied that by the corresponding pack size. We did this because many wolf pack territories overlapped more than one wolf management zone. In doing so, we effectively assigned a proportion of individual wolves in each pack  to each zone. Wolf range in each zone was defined as the amount of wolf range (5 km$^2$ buffer around wolf pack territories) occupied in each management zone.

We fitted logistic curves to population time series in each zone. Fitting these curves, enabled us to estimate carrying capacity for each zone. To validate these estimates, we compared them with carrying capacity estimates derived from species distribution models \citep{mladenoff2009change, stauffer2021scaling, gantchoff2022distribution}. For this comparison, we multiplied the average local density estimated above by the estimated suitable or occupied habitat of the species according to each distribution model.

To further understand spatial dynamics, we evaluated wolf range at similar spatiotemporal landmarks. Specifically, we identified years with zero-, second-, and third-order derivatives (i.e., inflection points) of each zone-specific fitted logistic growth curve. These points mark periods corresponding with early growth acceleration, maximum growth, and later deceleration of population growth, respectively. Additionally, we evaluated habitat boundaries modeled by \citet{mladenoff2009change} and \citet{ gantchoff2022distribution}. To further validate any patterns observed in Wisconsin, we also evaluated population and range growth patterns relative to habitat boundaries for the state of Michigan. We were able to do this because the data provided by the state of Michigan contained wolf population and range information. By comparing growth across spatial scales, we aimed to assess the consistency or divergence between the population dynamics and habitat boundaries at specific key points in space and time.

%% file: tex/results.tex

\subsection{Overall carrying capacity}

\begin{figure}[htp]
    \centering
    \includegraphics[width = .9\linewidth]{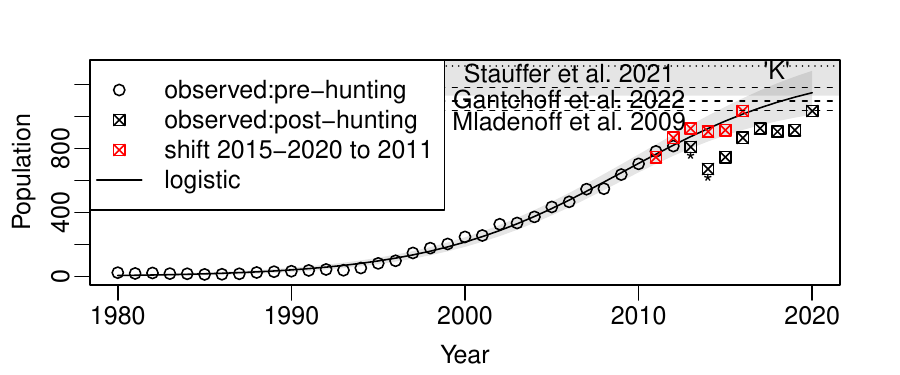}
    \includegraphics[width = .9\linewidth]{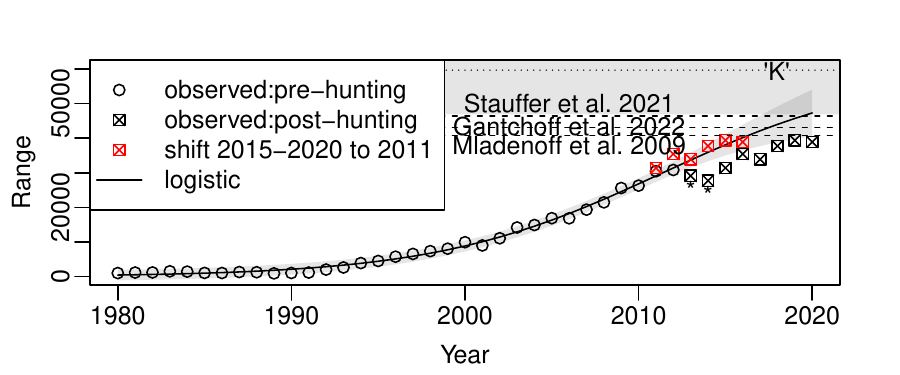}
    \caption{Growth and expansion of Wisconsin's wolf population (1980-2020). Logistic models were fit using data pre-hunting (prior to 2012). The likelihood ratio test indicates overall population growth and range expansion until 2012 were logistic than rather than exponential ($p < 1 \times 10^{-22}$). The logistic curve predicted shifted dynamics with root mean square error of 38.9, only 4.3\% of the mean predicted population size (898.3) for 2015-2020. Estimated carrying capacities derived the area of suitable wolf habitat \citet{mladenoff2009change, stauffer2021scaling} or occupied wolf range by \citet{gantchoff2022distribution} multiplied by 0.0254, the estimated stationary local density. The logistic model for range can predict the same shifted dynamic with root mean square error being 2,163.53, only 6.0\% of the predicted mean range (36,142.1) for 2015-2020. Estimates of suitable habitat from \cite{mladenoff2009change, stauffer2021scaling,gantchoff2022distribution} included for reference. }
    \label{fig:population_growth}
\end{figure}

Landscape-scale analysis revealed significant density dependence in both wolf range and total population size ($p < 10^{-22}$, Fig. \ref{fig:population_growth}). We estimated a state-wide carrying capacity of 1,315 (95\% CI = [1,131, 1,530]) for Wisconsin and a range limit of 59,696.44 km$^2$ (95\% CI = [45,453.9, 78,401.75]).

Our logistic model successfully predicted the population as if growth had been reset to the year prior to the harvest, with minimal error. The root mean square error (RMSE) for the population prediction was 38.9 wolves, representing only 4.3\% of average population size (Fig. \ref{fig:population_growth}). This indicates that our model exhibited strong predictive capability in terms of growth. Similarly, our range model also predicted wolf range following harvest. The RMSE for the range prediction was 2163.53 km$^2$, which accounted for 6.0\% of the average range (Fig. \ref{fig:population_growth}). These results further confirm our models' predictive accuracy in estimating the population dynamics and range expansion of wolves in Wisconsin, USA. 

\subsection{Average local density}

Regression on time ($p=0.13$) and KPSS test ($p>0.1$)  from 2000-2020 demonstrated that density was stable over the last two decades. Linear regression between range and population size revealed that wolf population size between 1980 and 2020 was explained well by changes in wolf range alone, without an intercept term. In fact, range alone accounted for 99\% of variance in population size. This indicates that wolf population size was proportional to area, supporting inference of a stable local density. The estimated local density based on this analysis was 25.4 individuals per 1,000 km$^2$ (95\% CI = [24.8, 25.9]).

Using our estimated stable local density and \citet{gantchoff2022distribution}'s habitat suitability estimates, we estimated population carrying capacity to be 1,545 (95\% CI = [1,176, 2,029]). This estimate aligns with our demographic-based estimate and agrees well with previous studies that employed both demographic-based approaches \citep{van2009growth} and habitat-based estimates (Fig.\ref{fig:population_growth}). 

\begin{figure}[h]
    \centering
    \includegraphics[width=.9\linewidth]{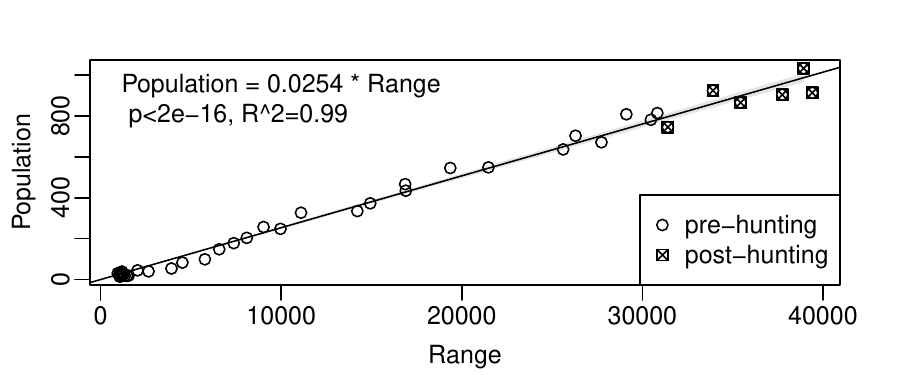}
    \includegraphics[width=.9\linewidth]{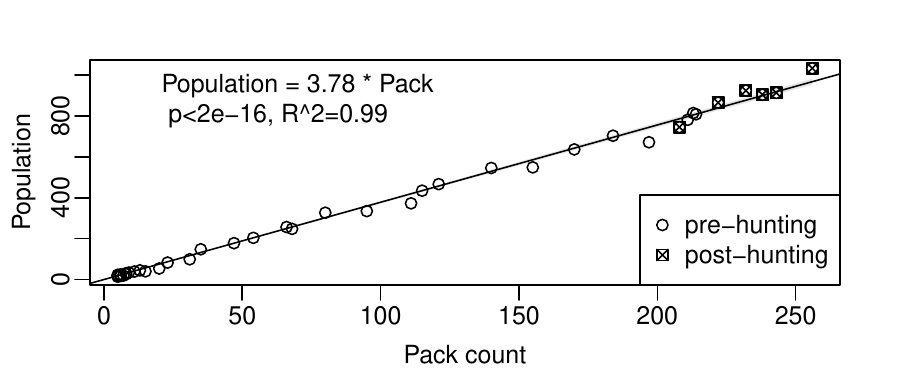}
    \caption{Range and pack count explained 99\% of variance in total wolf population (Wisconsin, USA, 1980-2020), suggesting support for a hypothesis of a traveling wave of the population and stable pack size. The estimated stationary local density is 25.4 individuals/1,000km$^2$ close to \citet{fuller2003wolf}.}
    \label{fig:expansion_driven}
    \label{fig:average_pack}
\end{figure}

\subsection{Pack and territory size}

Pack size alone (without intercept) explained 99\% of the variance in the total population, indicating stable average pack size estimated to be 3.78 individuals/pack (Fig.\ref{fig:average_pack}). The number was also stationary in 2000-2020, coincident with previous estimates in Wisconsin \citep{wydeven2009history}. Pack counts (without intercept) explained 99\% of variance in total range, indicating stable average wolf pack territory size estimated to be 148.6 km$^2$ (Fig.\ref{fig:average_territory}). The number was also stationary in 2000-2020, coincident with previous estimates in Wisconsin \citep{wydeven2009history}. This result was expected given results of population-range and population-pack count relationships and informs our exploration of potential local mechanisms.

\subsection{Local density and prey density}
\begin{figure}
    \centering
    \includegraphics{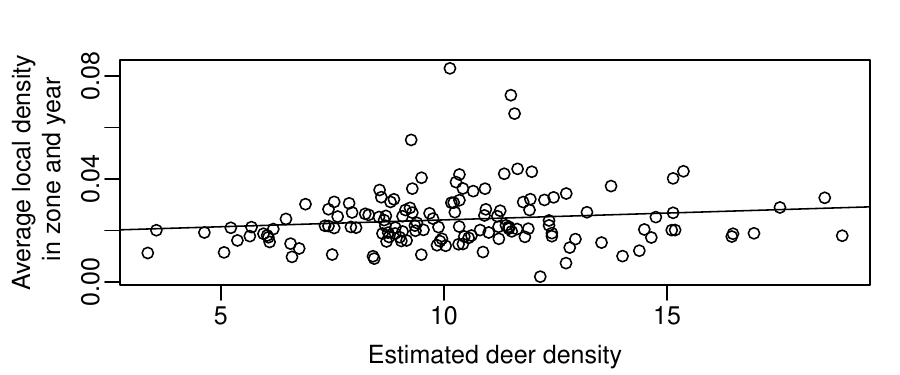}
    \caption{Deer density does not explain wolf density (all measured in individuals/km$^2$) in different zones and years. Some large local densities are from early years where measurement errors can be large.}
    \label{fig:deer_wolf_density}
\end{figure}

We found no significant linear relationship between deer densities and wolf local densities at the wolf management zone-scale ($p=0.08$, $R^2=0.015$, Fig.\ref{fig:deer_wolf_density}). 

\subsection{Mapping the expansion of wolves across spatial scales}

\begin{figure}[htp]
    \centering
    \includegraphics[width = \linewidth]{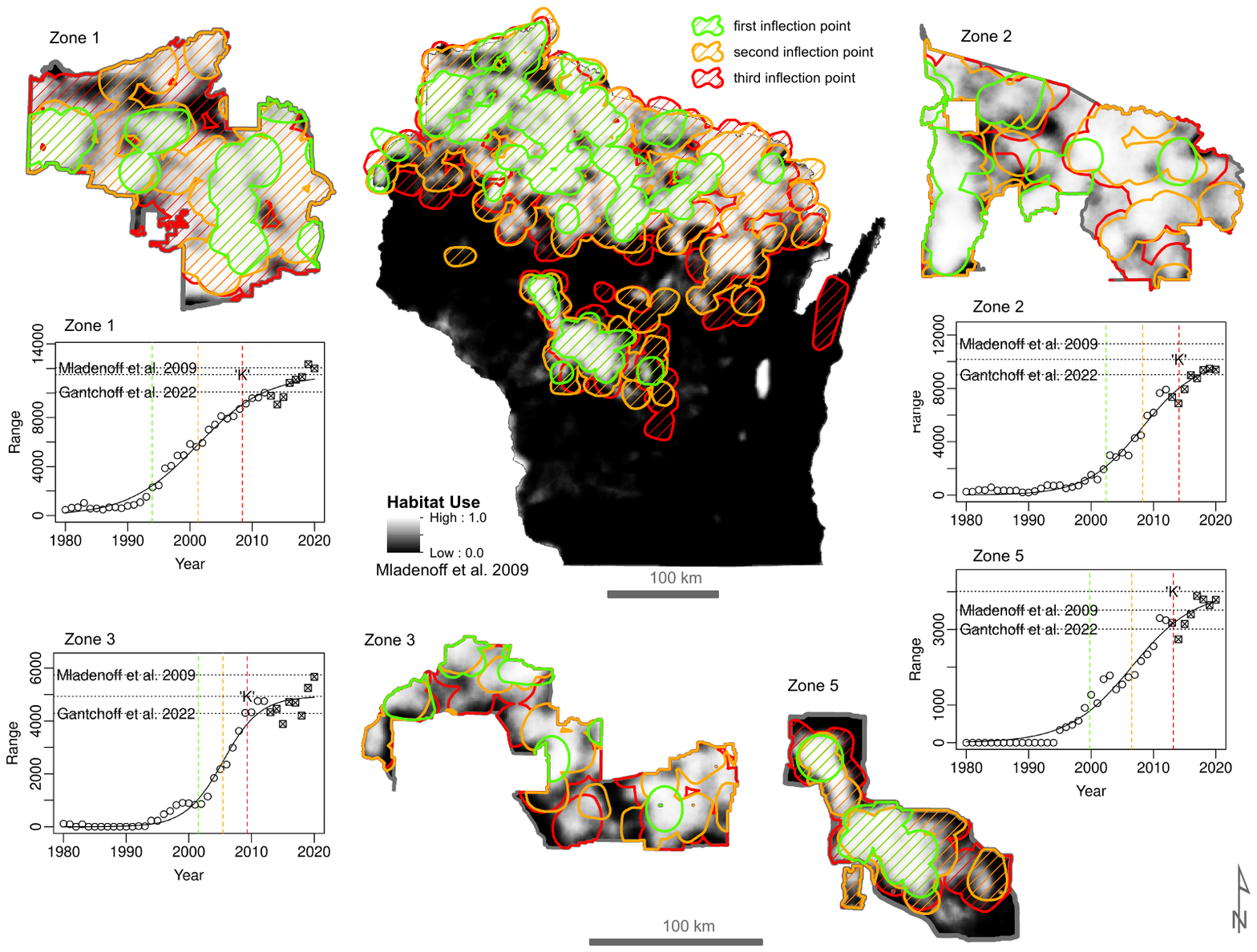}
    \caption{Mapping the expansion of wolves in Wisconsin in three demographic inflection points (when the curvature of the growth curve changes Green: first inflection, orange: second, red: third). Wolf ranges for years associated with inflection points from logistic growth curves for wolf range were mapped for the state and each zone and colored according to the corresponding inflection point. The first inflection point captures (green) when the frontier has established in the core of the habitat and expansion rates begin to accelerate. The second inflection point (orange) captures when the frontier has started to interact with the habitat boundary causing expansion rates to slow. The third inflection point (red) represents when the frontier has pushed into the boundary and expansion has begun to slow dramatically. The frontier generally expanded from the northwest corner of the state, where wolves first recolonized the state from Minnesota, towards the southern and eastern portions of suitable habitat in the state.  The effect of this pattern of recolonization can be observed in the shifting timing of inflection points for each zone.}
    \label{fig:wolf_map}
\end{figure}

Occupied wolf range in Wisconsin reflected habitat suitability mapping by \citet{mladenoff2009change,gantchoff2022distribution} and occupied wolf range as estimated by \citet{stauffer2021scaling} (Fig.\ref{fig:wolf_map}). Acceleration of wolf population growth (first second-order inflection point) at the state-scale coincided with the time when the frontier of wolf range was in the interior of the mapped suitable habitat. Later inflection points (i.e., maximum and decelerating growth) were achieved when wolf range frontier met the boundary of suitable habitat. Similar patterns occurred at the wolf management-zone scale (Fig.\ref{fig:wolf_map}). 

At wolf management zone-scale, we observed similar logistic-like growth curves for wolf population and range (Fig.\ref{fig:wolf_map}). In larger wolf management zones (e.g., Zones 1, 2, 3, \& 5), range-population relationships also exhibited similar patterns to what was observed at the statewide scale (Fig.\ref{fig:wolf_map}, \ref{fig:zone_level_linear}, \ref{fig:wi_mi_range}).
 
Notably, years associated with timing of first second-order inflection point, marking a period of growth acceleration, increased from the northwestern part of the state to the eastern portion of the state (Zone 1=1994, Zone 2=2002, Zone 3=2001, Zone 5=2000).   Since wolves colonized the state of Wisconsin initially in Zone 1 in northwestern Wisconsin, the timing of these first second-order inflections points captures the time-delay of the expanding frontier as it moved across zone boundaries.

%% file: tex/theory.tex

\subsection{The information puzzle}

Empirical results demonstrated ensemble density feedbacks at both landscape and local scales. Superficially, one may consider space as yet another limiting resource to explain landscape-scale density dependence. However, this view poses an information puzzle: how do territorial animals with limited perception ranges \citep{stenglein2015individual} ``know'' that their population's range is beginning to saturate suitable habitat to allow feedback on growth to operate? Do they need to know? Or does interaction between the expanding wave-front and habitat boundary make such information unnecessary?

\subsection{Frontier meets boundary: Fisher's reaction-diffusion model }

To address the central information puzzle and gain insights into the dynamics of population growth and range expansion, a simple dynamic system incorporating local density dependence and dispersal can be constructed using a reaction-diffusion model. Such a model can provide a framework to explore the interplay between population dynamics, range expansion, and habitat boundaries.

In this context, we reintroduce a two-dimensional, landscape-explicit version of the Fisher-KPP equation, originally introduced independently by Fisher and Kolmogorov, Petrovsky, and Piskunov (KPP) \citep{fisher1937wave, kolmogorovstudy}. This equation describes a reaction-diffusion process, capturing dynamics of population growth and dispersal. Our landscape-explicit version describes the interaction between an expanding wave front and a habitat boundary. Incorporating local density-dependent growth and dispersal mechanisms, this model can generate an expanding wolf range with stationary local density inside, resulting in a linear relationship between area and population. This phenomenon, referred to as a ``traveling wave'' solution, was initially observed in Fisher's original work on a unidimensional uniform landscape.

Further, presence of a boundary of suitable habitat that does not align perfectly with the expanding frontier can influence population dynamics. The interaction between the habitat boundary and the expanding frontier can give rise to a logistic-like growth curve, as the population approaches the limit imposed by available habitat. Using this simplified model, we can improve our understanding of the roles of range expansion, local density dependence, and habitat boundaries on population dynamics.

We started with Fisher-KPP equation, a partial differential equation with logistic local growth and a diffusion term
\begin{equation}
    \label{eq:fisher_kpp}
    \frac{\partial D(x,y)}{\partial t}=r_d\nabla^2 D(x,y) + r_g D(x,y)\left(1-\frac{D(x,y)}{K(x,y)}\right)
\end{equation}
where $D(x,y)$ is the local density landscape, $r_d$ and $r_g$ are rates for dispersal and local density growth, while $K(x,y)$ is the carrying capacity landscape. This landscape operates locally and could be based on prey regulation, human influence or other local mechanisms and we explore mechanisms in later sections. We denote $K_m$ as the maximum carrying capacity on the landscape and use it as the population unit while using $1/r_g$ as the time unit. We use the corresponding unitless transformation $D':=D/K_m$, $K'(x,y):=K(x,y)/K_m$ and $t':=r_gt$ $r':=r_d/r_g$. We interpret $r'$ as distance expanded in the time that the population was multiplied by $\exp(1)$, i.e. the time scale difference between dispersal and local growth. We simplified the model to its dimensionless form
\begin{equation}
\label{eq:simplified_PDE}
    \frac{\partial D'(x,y)}{\partial t'}=r'\nabla^2 D'(x,y) + D'(x,y)\left(1-\frac{D'(x,y)}{K'(x,y)}\right)
\end{equation}
Two important parameters determine behavior of the hypothetical population when the spatial unit is fixed: 1) time scale difference between local growth and dispersal $r'$ and 2) landscape (of carrying capacity) $K(x,y)$.

\begin{figure}[htp]
    \centering
    \includegraphics[width = \linewidth]{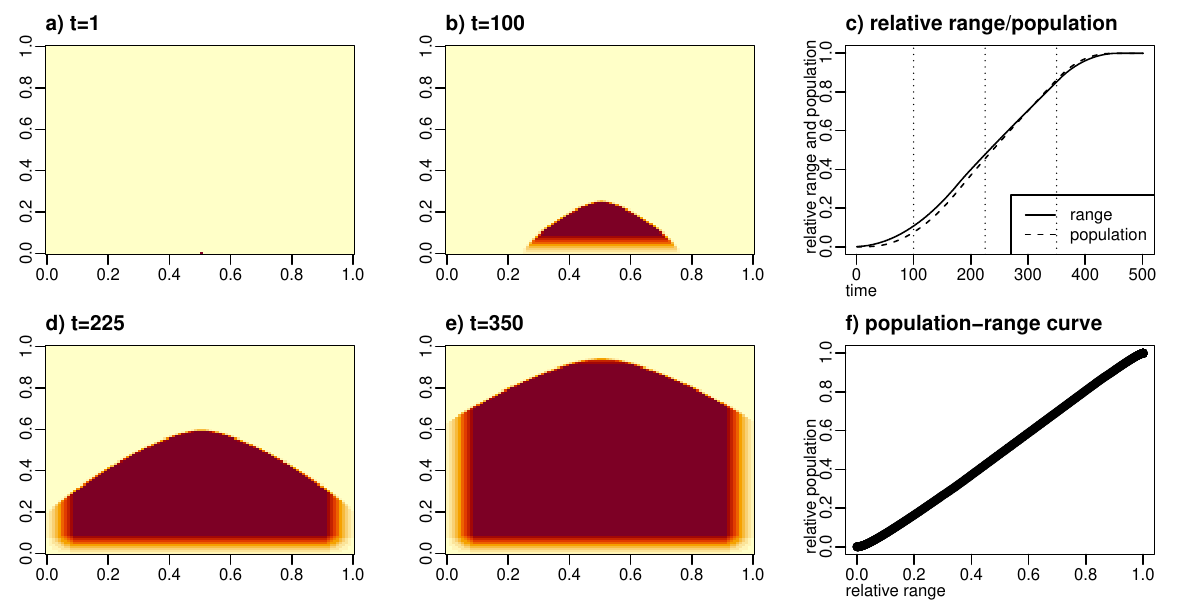}
    \caption{Fisher-KPP model with a ``soft'' habitat boundary. a), b), d), e): local density at four different time steps. c) the overall relative population and range showed a logistic-shaped curve f) the population-range curve is linear. 
    }
    \label{fig:softbound}
\end{figure}

We illustrated the numeric solution of model~\eqref{eq:simplified_PDE} in the scheme of $r'=r_d/r_g=0.02$ for 500 time steps. Using a range of $r'$ from 0.1 to 0.001 did not qualitatively change the result. At $r'\ge 1$, corresponding to a population that grows more slowly than dispersal, local extinction occurs under this model. We showed here the ``soft'' habitat boundary scheme, i.e.  where habitat suitability degrades with distance from core habitat (Fig.\ref{fig:softbound}) before carrying capacity hits 0 outside of the core habitat.  
We identified the frontier as spatial locations where local density is growing not because of dispersal but because local reproduction (i.e. non--zero second term). 

We observed a logistic-like growth curve  with a signature acceleration period in the initial stage of expansion when the frontier elongates and accelerates the growth process. As the frontier approaches and reaches the habitat boundary, expansion slows until it eventually stops. In the later stages of expansion, frontier length decreases as the population saturates suitable habitat within the habitat boundary, resulting in cessation of expansion.

Yet, habitat quality is rarely uniform, thus, it is also important to assess such patterns in more complex landscapes. Using our approach (above), we simulated population growth using a more complex landscape. We observed similar frontier-habitat boundary interactions and corresponding population growth in a more realistic landscapes, as depicted in Fig.\ref{fig:patchy} (Appendix \ref{sec:overall_growth_Fisher}). These visual and mathematical analyses demonstrate dynamic interaction between frontier and habitat boundary. Acceleration and deceleration of growth, as well as the eventual cessation of expansion, result from the interplay between expansion of the frontier and its eventual interaction with the habitat boundary.


\subsection{Geometric nature of the density dependence}

\begin{figure}[htp]
    \centering
    \includegraphics[width = \linewidth]{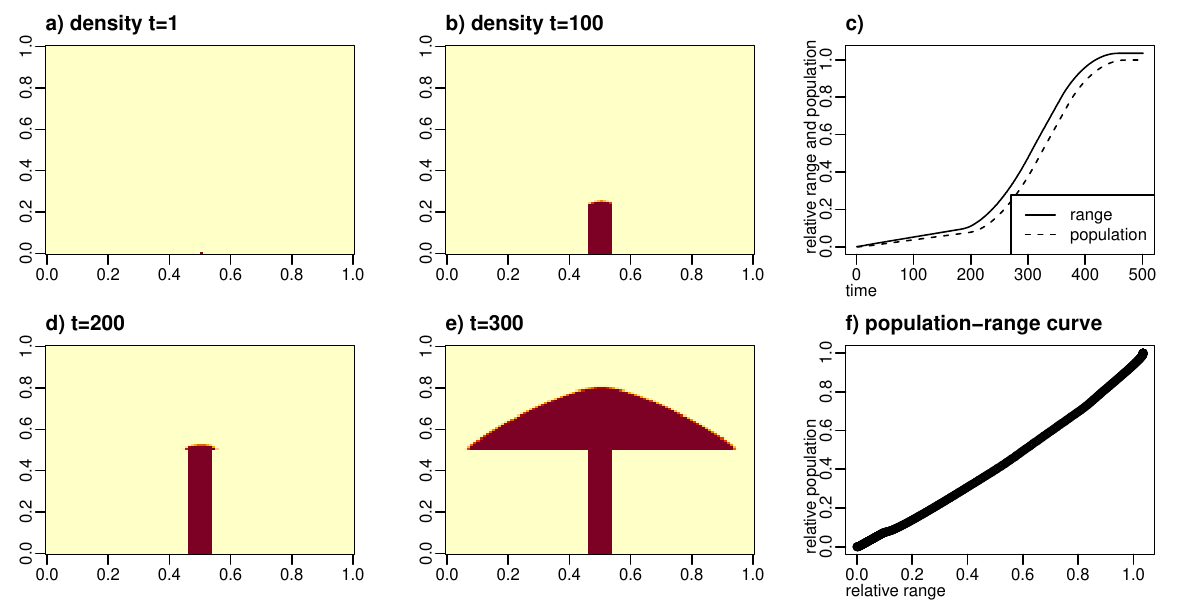}
    \caption{Reaction diffusion model with a population passing through a habitat corridor at the start of the timeseries can result in linear range-population relationship and Allee effect-like growth. a), b), d), e): Local density at four different time steps. c) Overall relative population and range, exhibiting an Allee-effect curve. f) Population-range relationship remains linear.  
    }
    \label{fig:allee}
\end{figure}

The traveling wave solution of Fisher-KPP equation does not depend on the form of local density dependence (second term in Equation~\eqref{eq:simplified_PDE}), instead, so long as quickly achieved local equilibrium and a dispersal term exist, a traveling wave could appear. In a uniform habitat, the dispersal frontier will be an expanding circle due to the symmetry of the dispersal term that does not prefer any particular direction. However, in reality habitat boundaries would not have a uniform geometry so an expanding frontier will hit a boundary at different times. The variable timing of frontier-boundary collision in complex or more realistic systems leads to a more exaggerated growth acceleration and deceleration. A quantitative prediction on the early stage growth in uniform habitat is that the growth curve should look more quadratic than exponential (e.g. in a logistic growth). This is because the traveling wave frontier will expand circularly at a constant rate and the area of such a circle is quadratic in its radius and time. Thus, the geometry of a habitat boundary can also lead to deviation from a logistic like curve. For instance, when the population has to pass through a corridor to reach a large habitat, the global growth curve could appear to have an Allee effect evidenced by an extended period of reduced growth at low density (Fig.\ref{fig:allee}).

\subsection{Traveling wave model on a realistic landscape: empirics}

We checked demographic curves with range maps of wolves at both the statewide- and zone-scales (Fig.~\ref{fig:wolf_map}), and found that population growth landmarks(evaluated by inflection points) aligned well with the timing of when frontiers were in interiors of mapped habitat and, later, clashing with habitat boundaries. This alignment between timing of inflection points in the demographic curves and the spatial dynamics of the expansion frontiers and habitat boundaries provides further support for our hypothesis that the interaction between population expansion and habitat boundary plays a crucial role in shaping the population dynamics of wolves in Wisconsin.

\begin{figure}[htp]
    \centering
    \includegraphics[width = \linewidth]{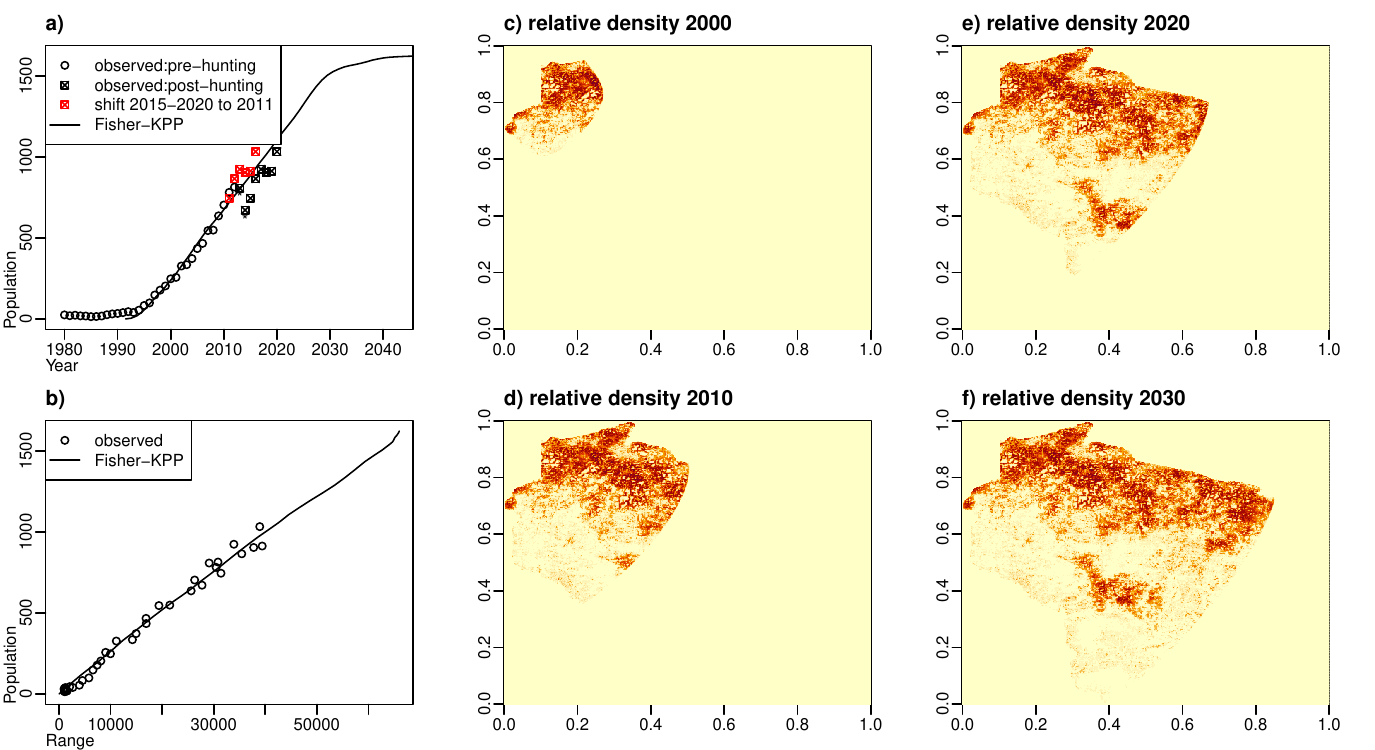}
    \caption{Fisher-KPP model simulated on Wisconsin (USA)'s wolf habitat map by \citet{gantchoff2022distribution}, fitted to observed growth curve. Comparison of observed a) population size and b) range-population relationships relative to Fisher-KPP model-simulation outputs. c)-f) Decadal relative wolf density according to our Fisher-KPP model and the wolf habitat predicted by Gantchoff et al. (2022).}
    \label{fig:wi_kpp_fit}
\end{figure}

Directly fitting the Fisher-KPP model to a real landscape is challenging. We developed a simplified procedure by fitting the linear population-range relationship and population growth curve but not the distribution. First, we used \citet{gantchoff2022distribution}'s habitat suitability map (normalized by its maximum) as the known $K'(x,y)$ and fitted local density $24.5/1,000$km$^2$ as $K_m$ (so we have $K(x,y)$ in equation~\eqref{eq:fisher_kpp}). The remaining task was to find $r_d$, $r_g$, or equivalently, $r'$ and time scale (the ratio between model time and real time in year) and potentially a time offset. We took the latter approach of determining the $r'$ first. To do so, We performed a grid search on the time scale separation $r'$, for each $r'$, we then fitted the time scale and a time offset via least square using the population growth curve as data. Finally, we showed the fitted growth curve, range-population relationship, and the predicted range at decades (Fig.\ref{fig:wi_kpp_fit}). This method does not use any wolf distribution data besides an initial value at northwestern corner of the state.  Our model, using a realistic habitat boundary, produced growth that resembled the observed range expansion for Wisconsin (Fig.\ref{fig:wolf_map}), further demonstrating the plausibility of the traveling wave model and its hypothetical mechanisms.

%% file: tex/niche_theory.tex

To better explore the concept and nature of the habitat boundary for wolves, we examined why carrying capacity $K(x,y)$ varies by location. Niche theory suggests  that each location possesses a unique environment, which can include factors such as elevation, topographical aspect, density of competitors, prey density, vegetation type, or other ecological variables \citep{hutchinson1957concluding}. Carrying capacity, therefore, may depend on the environmental characteristics of each location. Although the traveling wave behavior of Fisher's model does not hinge on how local carrying capacity operates, the determination of such carrying capacity is an important question otherwise.

\begin{figure}[htp]
    \centering
    \includegraphics[width = \linewidth]{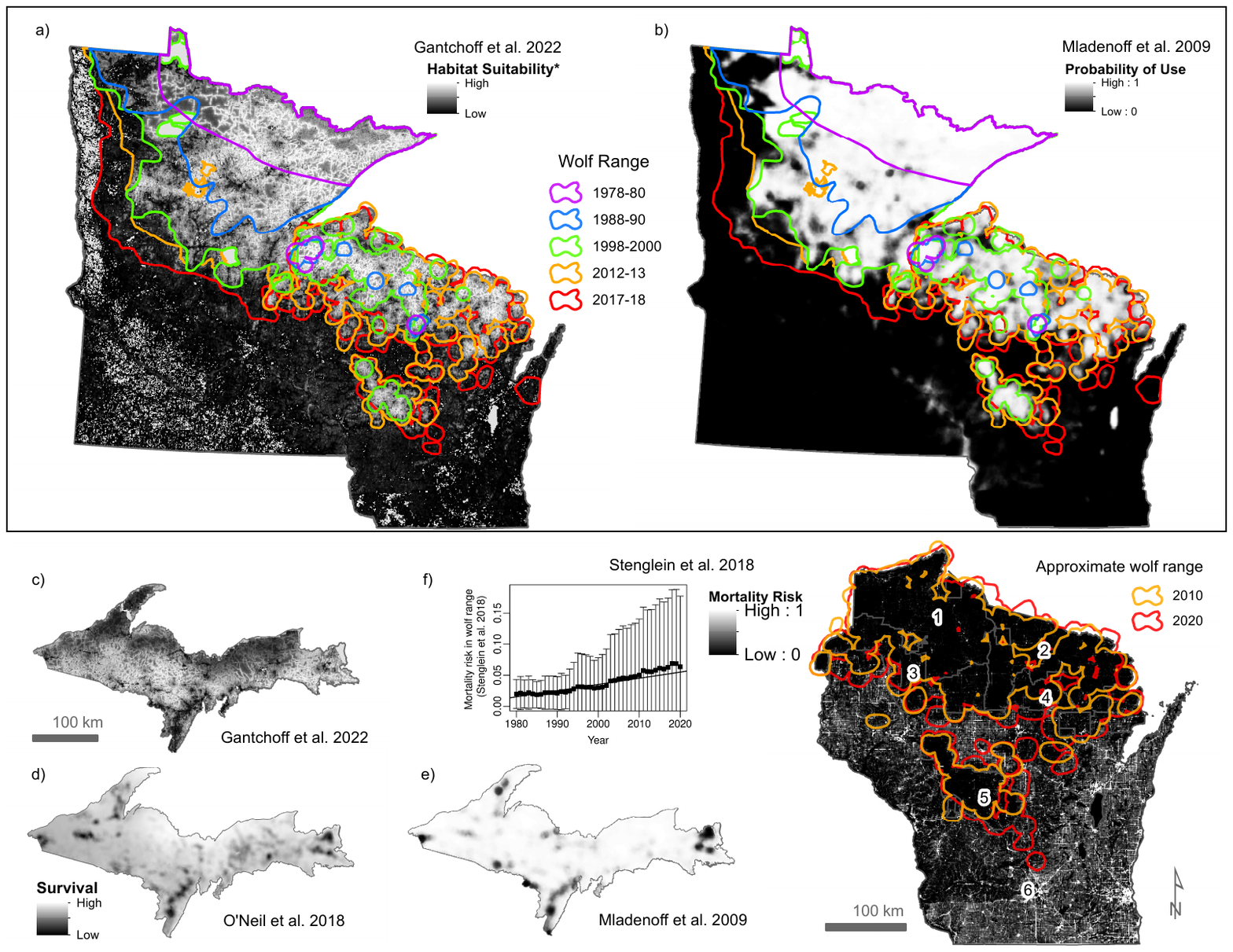}
    \caption{Expansion of wolves in Michigan, Minnesota, and Wisconsin (USA) relative to habitat boundaries and wolf mortality risk. a) Wolf habitat suitability according to \citet{gantchoff2022distribution} relative to wolf range over time for Minnesota and Wisconsin. Minnesota and Wisconsin's wolf ranges were developed using different methods, but are colored according to the time period associated with the wolf range. b) Wolf habitat suitability according to \citet{mladenoff2009change} relative to wolf range over time for Minnesota and Wisconsin. Michigan wolf habitat suitability according to c) \citet{gantchoff2022distribution} and d) \citet{mladenoff2009change}. e) Michigan wolf survival modeled by \citet{o2017spatially}. f) Wisconsin wolf mortality modeled by \citet{stenglein2018compensatory} relative to the Wisconsin wolf range in 2012-13 and 2017-18, and change over time in the mean and standard deviation of wolf mortality risk within the Wisconsin wolf range.}
    \label{fig:wolf_map_survival_gratelake}
\end{figure}

\subsection{Anthropological factors in determining habitat boundary: A synthesis of evidence}
According to niche theory, environmental variables establish a species' habitat boundary reflecting both fundamental and realized niches. In other words, a species in a particular habitat may have one carrying capacity landscape when free from competitors, and another when in the presence of those competitors A common finding among various habitat suitability studies was that human land-use is among the strongest predictors of whether a location is suitable for wolves \citep{mladenoff2009change,stenglein2018compensatory,gantchoff2022distribution}. For example \citet{mladenoff2009change} determined that increasing road density and agricultural land representation were negatively correlated with wolf pack distribution in Wisconsin. 

Historically, however, before European colonization of the USA, wolves occupied all terrestrial habitats present in Wisconsin and the western Great Lakes region so long as there was adequate prey \citep{beyer2009wolf,erb2009overview}. That is, because the fundamental niche of wolves would have allowed them to exploit essentially the entire region--so long as prey was present. Therefore, pre-settlement habitat boundaries for wolves would likely have been established by Great Lakes shorelines. Thus, one might ask -- What factors actually determine the contemporary habitat boundary (i.e., carrying capacity landscape) for wolves?

Researchers in Wisconsin and Michigan mapped wolf survival and mortality \citet{o2017spatially,stenglein2018compensatory}. Such spatial information yields valuable insights into habitat boundaries because a species will struggle to persist in areas where it has a greater risk of mortality. Thus, to better understand the nature of habitat boundaries for wolves in the western Great Lakes Region, we mapped regional patterns of wolf population growth relative to habitat boundaries or spatial patterns of wolf survival for Michigan, Minnesota, and Wisconsin (USA).

Due to differences in datasets available across the three states, we first presented wolf range maps for Minnesota in conjunction with our wolf range maps for Wisconsin from 1979-2018 relative to habitat suitability as estimated \citet{mladenoff2009change} and \citet{gantchoff2022distribution}. Separately, for the state of Michigan, we produced maps of wolf habitat suitability and wolf survival estimates from \citet{o2017spatially}. Similarly, for Wisconsin, we mapped wolf mortality probability as estimated by \citet{stenglein2018compensatory} and we evaluated change in mortality risk as wolf range expanded. This allowed us to examine visually the relationships between potential habitat boundaries, wolf range expansion over time, or wolf mortality or survival at a regional scale.

While the fundamental niche for wolves likely covered all three states, current wolf distribution in the region suggests that the realized niche of wolves' post-European colonization is reduced (Fig.\ref{fig:wolf_map_survival_gratelake}). Wolf range expansion in both Minnesota and Wisconsin correspond well with one another, despite coming from two separate data sources (Fig.\ref{fig:wolf_map_survival_gratelake}, a,b). The expansion of wolf range over time, also further reinforces the notion of an expanding frontier that initiated in northeastern Minnesota and expanded out from there until clashing with habitat boundaries, geopolitical borders aside (Fig.\ref{fig:wolf_map_survival_gratelake}, a,b). 

Wolf survival in Michigan \citep{o2017spatially} corresponds well with habitat suitability models (Fig.\ref{fig:wolf_map_survival_gratelake}, c, d, e). Wolf mortality probability in Wisconsin (Stenglein, Wydeven, and Van Deelen 2018) also corresponded well with habitat suitability models (Fig.\ref{fig:wolf_map_survival_gratelake}, a,b, f), suggesting that factors shaping wolf survival or mortality are fundamental in determining the habitat boundary. Human-caused mortality is an important mortality component for wolves in the western Great Lakes region. Vehicle collisions, accidental deaths, illegal killing, legal hunting and trapping, and lethal controls are all forms of  human-causes of mortality experienced by wolves in the region.  For example, in Michigan, human-caused mortality was the dominant source of mortality for wolves with approximately two-thirds of annual mortality attributed to human causes \citep{o2017spatially, midnr}. In Wisconsin from 1979 to 2012, human-caused morality accounted for approximately 60\% of annual mortality with greater risk of human-caused mortality being on the edges of suitable habitat \citep{stenglein2018compensatory}. 

Moreover, risk of mortality for wolves in Wisconsin increased over time as wolf range expanded (Fig. \ref{fig:wolf_map_survival_gratelake}, f), suggesting that as the wolf frontier in Wisconsin expanded, wolves established in areas with greater likelihood of mortality -- an ecological mechanism for the deceleration of growth as the frontier interacted with the boundary. The patterns observed suggest that post-European colonization, human population growth and land conversion, along with spatiotemporal variation in attitudes towards wolves, may have shaped the current realized niche of wolves in the region and concurrently their suitable habitat and range. I.e., humans have played an important role in determining the carrying capacity landscape for wolves in the region.

\subsection{Carrying capacity within habitat: prey vs territory}
In the traveling wave model, a stable local density occurs locally within the habitat boundary. We asked the question - What mechanisms reinforce such stable local densities? 
 
\begin{figure}[htp]
    \centering
    \includegraphics{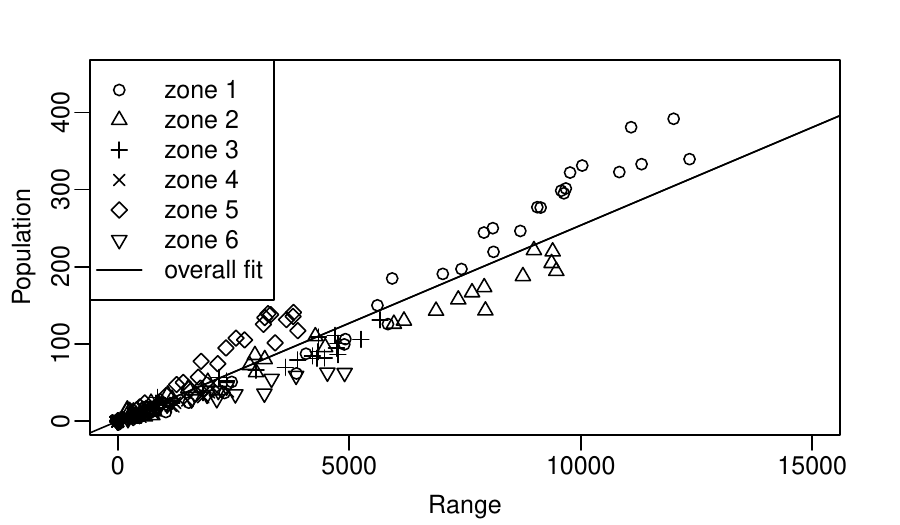}
    \caption{Range-population relationships for wolf management zones in Wisconsin, USA. Zones exhibit a similar linear pattern to the overall fit using aggregated data.}
    \label{fig:zone_level_linear}
\end{figure}

For large predators, prey availability is critical to survival and reproduction. If prey is the mechanism, variation in prey density throughout suitable habitat may influence space use as predators adjust territory size. In our study system, prey density can vary substantially (e.g., 1.2-19 deer/km$^2$, for the northern or central forests of Wisconsin(2014 to 2020; Wisconsin Department of Natural Resources), with white-tailed deer  being the dominant prey range-wide, \citep{delgiudice2009prey}). Such variation in prey density could explain variation in the range-population relationship at the zone scale (Fig.\ref{fig:zone_level_linear}). We also found that wolf pack territory size was relatively stable (Fig.\ref{fig:average_territory}, \ref{fig:deer_wolf_density}). Over time, recovering wolf populations in Wisconsin would have expanded into areas with more variable prey densities and prey density would have varied over time. However, despite variation in prey density over time and space we see relatively stable local and global densities and pack and territory sizes for wolves (Fig.\ref{fig:zone_level_linear}, \ref{fig:average_territory}), suggesting that prey may not be the main regulatory mechanism for wolves in Wisconsin.

%% file: tex/discussion.tex

\paragraph{Mechanism for large scale density dependence.} To understand integration of population and range dynamics at landscape scale, we studied the population-range relationship for a recovering wolf population in Wisconsin USA. Results indicated a near-constant local density of 25.4 individual/1,000km$^2$ and a linear population-range relationship.

The phenomenon of linear range-population relationship matches the ``traveling wave'' solution of Fisher-KPP model~\eqref{eq:simplified_PDE}. Logistic-like growth range can be reproduced by having habitat boundaries and the interaction between an expansion frontier and a habitat boundary, which was not possible in Fisher’s original unidimensional case where both boundary and frontiers are points with no length. When expansion happens in two dimensions, both frontier and boundary are curves while frontier has changing length. Change in length of frontier could change  overall growth rate. Alternatively, a reduction in dispersal at intermediate densities could be another mechanism driving logistic-like growth. A traveling wave process \citep{fisher1937wave,kolmogorovstudy} is a useful model for describing ensemble density dependence in the recovery of Great Lake wolf populations in large scale because it solves the ``information puzzle'' embedded in wolf life history -- that being the hypothetical need for information on the process of global saturation to inform component density feedbacks. Our analysis suggests that local density dependence and an expanding wave-front interacting with habitat boundaries is enough to reproduce spatio-temporal patterns of density dependence that reflect observed patterns. Moreover, consideration of habitat boundaries suggested that the key component density feedbacks are territorially and social constraints on pack size and spatially varying mortality risk associated with anthropogenic land-use (i.e., when suitable habitat is saturated, more wolves are forced into risker landscape areas).

Further analysis of wolfpack size indicated a near-constant average pack size of 3.78 individuals/pack (and a near-constant average wolf pack territory size). Relatively constant local density and pack size suggested that, at least for landscapes with adequate and relatively uniform prey composition, wolves in Wisconsin expand their range by forming new packs of similar sizes in terms of individuals and territory.
However, the Fisher-KPP model does not account for social behaviors of wolves. Further, the dispersal term in the model could be understood as Brownian behavior of dispersers with unimodal light-tailed dispersing distance (i.e. extremely long-distance travelers are very rare). However, the distance traveled by disperser wolves can be heavy-tailed (i.e. extremely long distance travelers can be less rare) or bimodal  \citep{mech2020unexplained,2004mech} and potentially, directional \citep{gable2019there}, thus some (extremely long distance traveling) individuals might establish relatively isolated or disjunct sub-populations \citep[Fig. \ref{fig:wolf_map_survival_gratelake} a-b, see also ][Ch.7, pg. 107-117]{wydeven2009recovery}. These new isolated subpopulations might establish new wavefronts speeding up the expansion. Thus, a more complicated model that accounts for group-living behaviors as well as the dispersal characteristics of wolves might provide further insights.

Additionally, fitting parameters in the Fisher-KPP model is challenging, especially when including a carrying capacity landscape that is unknown. We approximated the carrying capacity landscape using habitat suitability estimates from the literature. Another option would be to model carrying capacity based on environmental variables in a joint manner. However, fitting these models can be computationally demanding, as it often involves solving the equation repeatedly. The fitting method for partial differential equation models is an active research topic among statisticians and some approximated methods might be used, e.g. approximated Bayesian computing \citep[ABC,][]{csillery2010approximate} among other neural network based simulation based inference methods \citep[e.g., reviewed by][]{Cranmer2020}.

\paragraph{Habitat boundary and carrying capacity locally.} Our results suggest a mechanism of large scale density dependence requiring only local density information: the logistic-like curve can be mediated by the elongation of the frontier and its eventual interaction with a habitat boundary while the carrying capacity was determined by the total area and habitat quality. Synthesizing previous studies, we suspect that current observed habitat boundaries for wolves in the western Great Lakes region likely are established by spatial variation in mortality, especially human-caused mortality, which is expected to increase in areas with a greater human footprint on the landscape \citep{stenglein2018compensatory}.  Thus, future changes in human population, land cover, and attitudes towards wolves could lead to the expansion or contraction of the observed habitat boundary and in turn influence wolf population dynamics \citep{mech2017can}.

\citet{o2019territorial} demonstrated that as wolf densities in Michigan increased wolf territory overlap increased and habitat selection preferences shifted towards greater use of lower quality habitat.
Dispersal in wolves is the mechanism of range expansion \citep{harrington2003wolf, kojola2006dispersal, mech2017can}. Yet, when wolf range expansion clashes with a habitat boundary dictated by spatial variation in mortality, dispersing wolves that cross the boundary have higher mortality  \citep[Fig.\ref{fig:wolf_map_survival_gratelake},][]{o2017spatially, o2019territorial}. In Finland, dispersing wolves were less likely to avoid roads and human structures even after establishing a pack \citep{barry2020does}. Wolves dispersing outside of habitat boundaries are more likely to die from vehicle collisions because of the higher density of roads \citep{mladenoff2009change, stenglein2018compensatory,morales2022patterns} and to experience higher turnover rates (i.e., cyclical colonization and extinction) \citep[e.g.][]{mech2019gray, mech2017can}. Wolves or packs established outside of boundaries were also more likely to come into conflict with humans \citet{olson2015characterizing,mech2019gray, olson2019landscape,barry2020does}, which could consequentially trigger higher human-caused mortality, legally or illegally. According to \citet{olson2015pendulum}, over time, as wolf populations grew, wolf-human conflict became increasingly more common, even relative to wolf population size or range. \citet{mech2019gray} documented such an event when a pack that established outside of suitable wolf habitat began repeatedly depredating on domestic animals, and resulted in the state implementing lethal control (killing) to remove the pack. In another example, \citet{kojola2006dispersal} documented that wolves dispersing north into reindeer management areas in Finland were killed before they were able to reproduce. As wolf range expansion collides with the human-dictated habitat boundary. Our work suggests that for contemporary landscapes,  humans have played an important role in determining the carrying capacity landscape for wolves in the region.

\paragraph{Self regulation for top carnivores}
The fact that essentially all resources, such as prey, are also spatial adds complications to such questions, since space, within their perception ranges, can also be an important resource - especially for territorial species. Further,  there is an important distinction between population limitation and regulation - resources can be limiting, but population regulation requires identification of density-dependent feedback mechanisms \citep{messier1994ungulate}. Prey density may be a limiting factor, but requires the assumption that territories are established and maintained on the basis of prey density alone. We suspect that when prey density reaches a certain threshold there may be a decoupling of predator-prey dynamics and other resources, like space, may become limiting with intraspecific strife regulating population growth via territorial maintenance. Such an inference is perhaps supported by the observed stable local and global densities and pack and territory sizes throughout time and space.

Self regulation via territorial behavior of wolves is a potential mechanism for the observed stable local densities within the habitat boundary. Territoriality is, essentially, a form of contest competition; whereby individuals or social groups compete for space (i.e., territory) to reduce risk of intraspecific strife over resources required for survival and reproduction \citep{wilson2000sociobiology,2004mech}. Such negative interactions are mediated through behavioral feedbacks (e.g., olfactory, visual, or auditory cues) that communicate with potential intraspecific competitors that, ``This space is occupied'' to minimize intraspecific strife  \citep{2004mech}.  

For wolves, territoriality is established and maintained through behavioral feedbacks associated with scent marking and vocalizations that produces a bowl-shaped distribution of scent-marks for a given territory, resulting in honeycomb-shaped territories \citep{peters1975scent, lewis1997analysis, 2004mech}. Such honeycomb-shaped territories are regularly observed in reality for wolf populations \citep[e.g.,][]{ lewis1997analysis}, suggesting these feedbacks to be relevant. Yet, more recent models allowing the effectiveness of scent marks to fade over time produced irregularly-shaped honeycomb territorial boundaries that shifted over time in response to the behaviors of individuals in neighboring territories \citep{giuggioli2011animal}. Recent use of GPS collars on wolves in the Great Lakes region supports these results; demonstrating a greater shared use of the interstitial space between packs and more irregularly-shaped honeycomb territories as wolves respond to both the landscape and neighboring packs \citep[][D. MacFarland, pers. comm.; see also www.voyageurswolfproject.org/animations, ]{Nordin2023assessing,gable2016and}. Such behavioral mechanisms underpin the landscape scale patterns of growth for both population and range as observed and modeled in our study. Locally, these behavioral feedbacks associated with scent marking, when applied to a recolonizing population, appear to translate into the only local information needed to exhibit the growth phase of the logistic-like growth patterns observed, with local density likely saturating quickly at high prey densities.

\paragraph{Conservation Implications.}
Our results suggest that persistence on the landscape is inherently a spatial phenomenon (i.e., a carrying capacity landscape). This has important policy implications. The federal status in the US for wolves under the Endangered Species Act in the western Great Lakes region has been unstable and prone to quick swings from federally Endangered to fully delisted and a state-designated game species \citep[e.g.][]{olson2015pendulum}. In \citet{defendersvusfw2022} in the US, a federal judge returned wolves back to the federal list of Endangered species on the basis that wolves had not yet returned to a “significant portion of range.” \citep{bruskotter2013predator}.  Habitat boundaries for wolves appears to be dictated by human activities and infrastructure and  wolves likely existed nearly everywhere in North America prior to European colonization. If the habitat carrying capacity (or realized niche) of wolves is established by social forces dictating human behavior and land-use -- how then do we determine what a ``significant portion of range'' represents? It is clear that suitable wolf habitat exists outside of currently occupied wolf range \citep{van2022identifying}. It is also clear that some of these areas appear to have acceptable levels of social tolerance for wolves currently \citep[][Colorado, USA]{bruskotter2013predator}. 

Our work adds another line of evidence suggesting that wolves exhibit negative density dependence in their population biology as seen in the numerical and spatial patterns of growth \citep{van2009growth,stenglein2016demographic} and in components of growth like survival \citep{cubaynes2014density, o2017spatially,stenglein2018compensatory}, reproduction \citep{stenglein2015numbers} and compensatory behavior in cause-specific mortalities \citep{stenglein2018compensatory}. 
This work also suggests that an important metric of wolf populations is the amount of area occupied by wolves. Multiple jurisdictions in the USA currently use patch occupancy models, spatial models of wolf pack occupancy, to estimate wolf populations \citep[e.g., ][]{miller2013determining,sells2022integrating,ausband2022estimating,moeller2018three,stauffer2021scaling}. Our work suggests that such approaches likely reflect a realistic mechanistic understanding of population growth and range expansion.

\paragraph{Future directions.} Our results suggest that seemingly non-spatial phenomena (e.g. logistic growth) could arise from spatial context and seemingly distinct phenomena (e.g. Allee effect and logistic growth) could arise from the same simple spatial geometric mechanisms. The distinction between a spatial mechanism and the classic perspective of logistic growth mechanisms lies in how feedback information informs growth. Our work suggests that spatial dynamics and dispersal mechanisms should be an integral part of population dynamics theory. 

Beyond single species, multiple species system might facing the same information problem in e.g. Lotka-Volterra models \citep[e.g.,][applied Lotka-Volterra model on landscape]{andrennumerical}. In such cases, the habitat boundary are no longer static but dynamically determined by interaction that induces a changing geometry. Whether these geometrical dynamics of boundaries can induce similar behaviors like Lotka-Volterra models in well-mixed systems is an interesting question to explore in the future.  

We aim to draw more insights on whether similar patterns can be observed in population and spatial dynamics of wolves and other large carnivores in distinct landscapes and geographic regions. A collaborative effort to compare data from different parts of the globe with recovering wolf populations would be vastly beneficial to our understanding of this species and density dependence across different scales

%% file: tex/acknowledgement.tex
We thank the MI, MN, and WI Department of Natural Resources staff responsible for wolf monitoring for the past four decades, which was, and still is, extremely challenging. Without them and the data generated by their efforts, this study would be impossible. We also thank authors of \citet{mladenoff2009change, o2017spatially, stenglein2018compensatory} and \citet{gantchoff2022distribution} for their willingness to share spatial data generated as a part of their respective research. Without their efforts and data sharing it would have been difficult to explore the nature of the frontier-boundary concept. We thank David McFarland (WIDNR) for his helpful comments in the early stage of this study. YS thanks  Mingzhang Liu, Hongliang Bu, Jiangyue Wang (PKU) and Binxu Wang (Harvard) for helpful discussions during theory development, as well as David Burt and Renato Berlinghieri (MIT) for discussions on methods to fit the model. ERO wishes to acknowledge the Northland College Sabbatical and Professional Development programs, the Raymond D. Peters Professorship in Biology from Northland College, and his family for their continued support. 

%% file: tex/appendix.tex
\appendix
\renewcommand{\theequation}{A\arabic{equation}}
\renewcommand{\thesection}{A\arabic{section}}  
\renewcommand{\thefigure}{A\arabic{figure}}  
\renewcommand{\thesubsection}{A\arabic{subsection}} 
\renewcommand{\thesubsubsection}{A\arabic{subsubsection}} 
\setcounter{equation}{0}
\setcounter{figure}{0}
\setcounter{section}{0}
\setcounter{subsection}{0}
\setcounter{subsubsection}{0}



\section{Pack territory}
\begin{figure}[H]
    \centering
    \includegraphics[width=.9\linewidth]{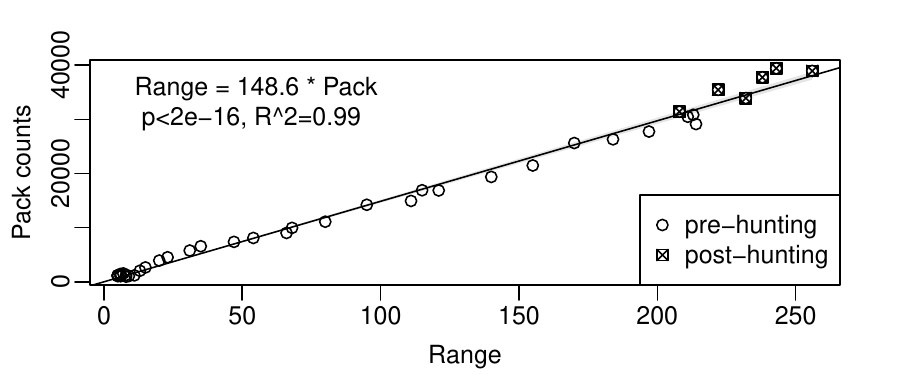}
    \caption{Pack count explained 99\% of variance in total wolf range (Wisconsin, USA, 1980-2020).}
    \label{fig:average_territory}
\end{figure}

\section{Zone level population-range relationship}

\section{Population-range relationship in WI and MI}
\begin{figure}[H]
    \centering
    \includegraphics{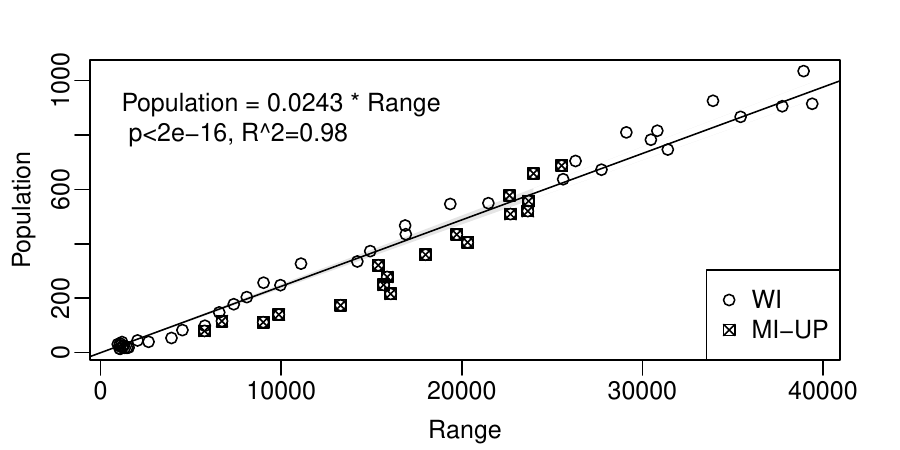}
    \caption{Range-population curve for WI and MI combined}
    \label{fig:wi_mi_range}
\end{figure}

\section{Other type of habitat boundaries}
In this appendix, we show the numeric solution of two different landscape types, namely a hard boundary habitat and a randomly generated habitat with an exponentially decayed correlation function.

\begin{figure}[H]
    \centering
    \includegraphics[width = \linewidth]{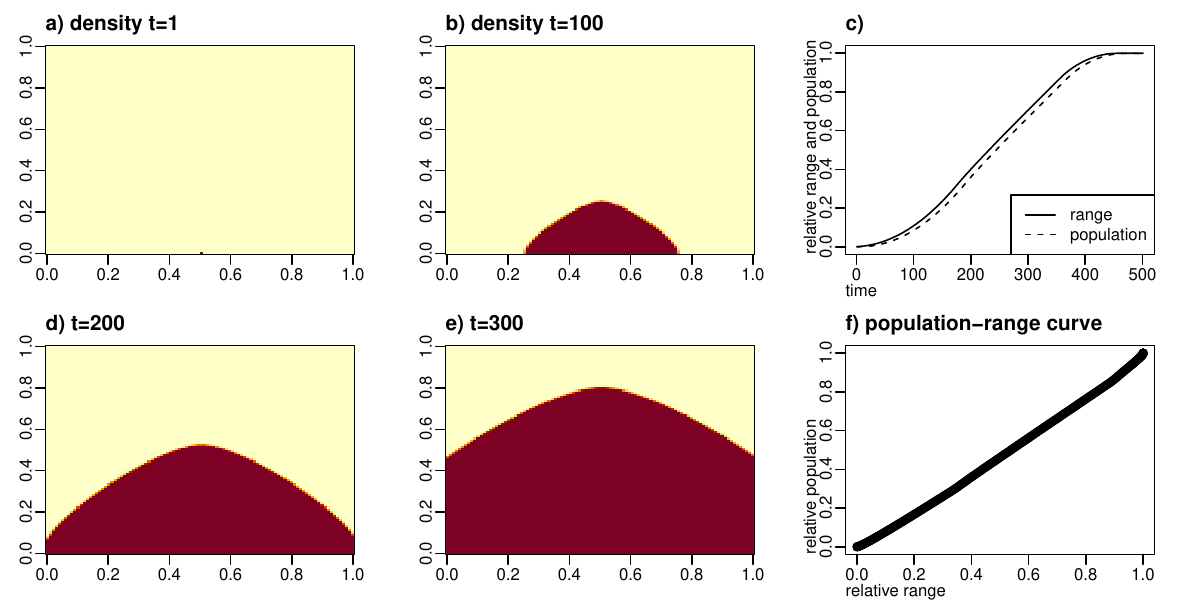}
    \caption{Reaction diffusion model with hard boundary. a), b), d), e): local density at four different time steps. c) the overall relative population and range showed a logistic-shaped curve f) the population-range curve is linear. 
    }
    \label{fig:hardboundary}
\end{figure}

\begin{figure}[H]
    \centering
    \includegraphics[width = \linewidth]{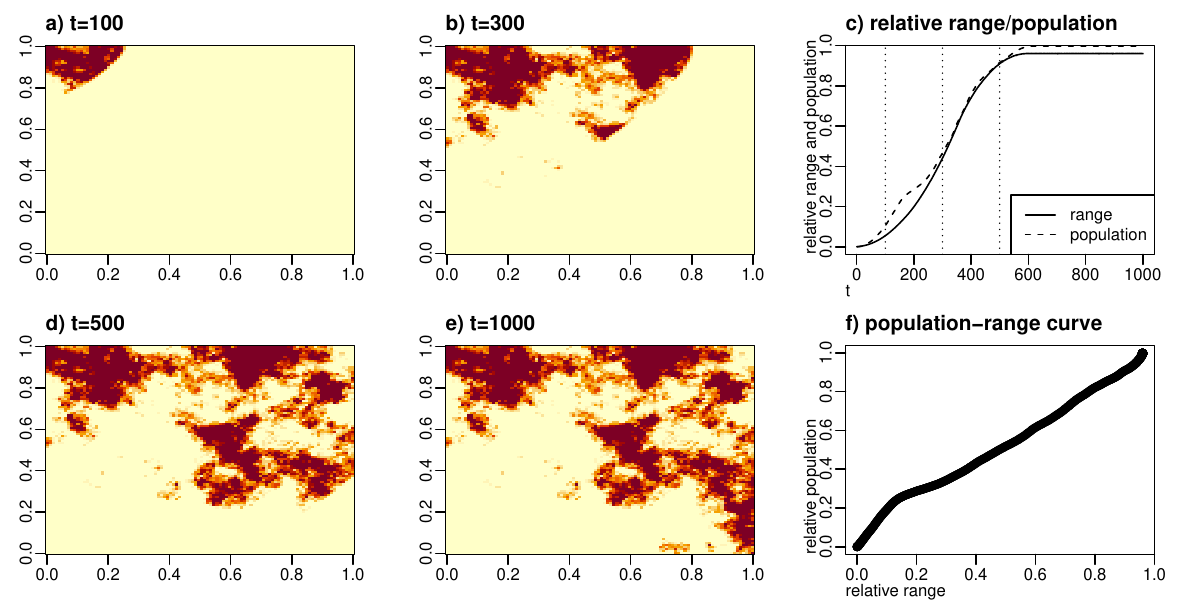}
    \caption{Reaction diffusion model with a realistic patchy habitat and time scale separation will result in linear range-population relationship and logistic-like growth. a), b), d), e): local density at four different time steps. c) the overall relative population and range showed a logistic-shaped curve f) the population-range curve is linear. 
    }
    \label{fig:patchy}
\end{figure}

\section{Mathematical analysis on the overall growth under Fisher-KPP model}
\label{sec:overall_growth_Fisher}

\subsection{On heterogeneous landscape: connecting local and overall densities}

There exists a rich body of literature on wolf population dynamics \citep{van2009growth,stenglein2016demographic,o2017spatially}, density-dependence \citep{sinclair1996density, herrando2012density,van2009growth,stenglein2016demographic}, and spatial usage \citep{van2009growth,o2017spatially,stenglein2016demographic}. The word \textit{density} is sometimes used in both local and global contexts. Since we work across different scales, we formalize some words to be used to avoid confusion. 

We start with the study area, or \textit{domain} $\cD$, a collection of spatial locations $(x,y)$ we are interested in and inside which the population and range were counted. We also denote the area to be $\vert\cD \vert$. Then we consider \textit{local density} $D(x,y)$, a function of spatial location, the population per unit area in that location $(x,y)$. Then the total population will be an integral of density $P=\int_\cD D(x,y)dxdy$. The \textit{overall density} is the average, i.e. total population divided by the total area of domain $P/\vert\cD \vert$. In a well-mixed population, i.e. individuals are free to be anywhere, the density $D(x,y)$ will be constant then both densities coincide. The distinction between local and over-all density is only important when the landscape is heterogeneous. 

Wolves achieve a well-known local carrying capacity. The landscape for wolves can be collected as a \textit{local carrying capacity} landscape $K(x,y)$, i.e. the maximum number of individuals per unit area at a certain location $(x,y)$. It is a landscape because it is a function of locations. Then the maximum allowed population will be the integral of local carrying capacity $K=\int_\cD K(x,y)dxdy$, then the \textit{overall carrying capacity} for density is $K/\vert \cD\vert$. Similarly, if $K(x,y)$ is a constant, then local and overall coincide. 

When we have these definitions, we can unify our view on some seemingly distinct spatial locations based on the value of $D(x,y)$ and $K(x,y)$. For instance, the \textit{range}, where there exist wolves, can be a collection of locations where local density is positive, and more importantly, the habitat $\cH$, the collection of locations with positive local carrying capacity, i.e. $\cH=\{(x,y): K(x,y)>0\}$. Species distribution models can be viewed as a classification task on finding locations in the habitat.

The total population can be calculated by integrating density across domain (i.e. the entire landscape, denote as $\cD$ whose boundary is $S$). Assume we can exchange derivative and integral in the traveling wave solution the growth of total population (a scaled version of the overall density) is then governed by an ODE
\begin{align}
\begin{split}
    \frac{d}{dt}\iint_{\cD} D dxdy&=\iint_{\cD}\frac{\partial D}{\partial t} dxdy\\
    &=r\iint_{\cD}\nabla\cdot\nabla Ddxdy + \iint_{\cD} D(1-\frac{D}{K(x,y)})dxdy\\
    &=r\oint \nabla D\cdot n dS+\iint_{\cD} D(1-\frac{D}{K(x,y)})dxdy\\
\end{split}
\end{align}

In the last line we used the divergence theorem. The first term correspond to the line integral of boundary of the study area (e.g. Wisconsin) and the flow out. This term is not very important when the domain was chosen so that it only passes through non-habitat or core habitat thus no gradient. The second term correspond to the growth due to local density growth across the entire landscape. In the traveling wave solution of Fisher-KPP equation, only two type of location will have non-0 second term: 1) the source-sink population around the habitat boundary where $K(x,y)$ is small but close to somewhere $K(x,y)$ is larger and 2) the frontier between the equilibrium of local density equals to carrying capacity and the equilibrium of no wolves will have the second term being non-0 and the overall growth will be dominated by the area within the frontier. We can partition the domain 
\begin{align}
\begin{split}
\iint_{\cD} D(1-\frac{D}{K(x,y)})dxdy=&\iint_{\text{source}} D(1-\frac{D}{K(x,y)})dxdy\\
&+ \iint_{\text{sink}} D(1-\frac{D}{K(x,y)})dxdy\\
&+ \iint_{\text{frontier}} D(1-\frac{D}{K(x,y)})dxdy
\end{split}
\end{align}
with
\begin{align*}
    \text{source}&=\left\{(x,y):D(x,y)<K(x,y), \nabla^2D<0,\frac{\partial}{\partial t}D(x,y)=0 \right\}\\
    \text{sink}&=\left\{(x,y):D(x,y)>K(x,y), \nabla^2D>0,\frac{\partial}{\partial t}D(x,y)=0 \right\}\\
    \text{frontier}&=\left\{(x,y):D(x,y)\ne K(x,y), \frac{\partial}{\partial t}D(x,y)\ne 0 \right\}\\
\end{align*}
In this formulation, the frontier can also be a ``shrinkage'' frontier if the initial value is larger than the carrying capacity. The crucial difference between source and frontier is whether it is at equilibrium and thus contributes to the overall population growth.